\documentclass[12pt]{article}
\usepackage{epsfig}

\def\gray{\special{ps: 0.4 setgray}}
\def\black{\special{ps: 0.0 setgray}}

\newcommand{\draft}{
\newcount\timecount
\newcount\hours \newcount\minutes  \newcount\temp \newcount\pmhours
 
\hours = \time
\divide\hours by 60
\temp = \hours
\multiply\temp by 60
\minutes = \time
\advance\minutes by -\temp
\def\hour{\the\hours}
\def\minute{\ifnum\minutes<10 0\the\minutes
            \else\the\minutes\fi}
\def\clock{
\ifnum\hours=0 12:\minute\ AM
\else\ifnum\hours<12 \hour:\minute\ AM
      \else\ifnum\hours=12 12:\minute\ PM
            \else\ifnum\hours>12
                 \pmhours=\hours
                 \advance\pmhours by -12
                 \the\pmhours:\minute\ PM
                 \fi
            \fi
      \fi
\fi
}
\def\fullclock{\hour:\minute}
\begin{centering}
\gray
\special{ps: -90 rotate}
\special{ps: -5100 -5000 translate}
\font\Hugett  =cmtt12 scaled\magstep3
{\Hugett Draft: \today,\clock}
\black
\special{ps: 90 rotate}
\special{ps: 5000 -5100 translate}
\end{centering}
\vskip -1.7cm
$\phantom{a}$
} 
\catcode`\@=11 
\def\lsim{\mathrel{\mathpalette\@versim<}}
\def\gsim{\mathrel{\mathpalette\@versim>}}
\def\@versim#1#2{\vcenter{\offinterlineskip
        \ialign{$\m@th#1\hfil##\hfil$\crcr#2\crcr\sim\crcr } }}
\catcode`\@=12 

\def\nextline{\hfill\break}

\def\mycomm#1{\nextline\strut\kern-3em{\tt ====> #1}\nextline}

\def\nextline{\hfill\break}

\newcommand{\ee}{\hbox{$e^+e^-$}}
\newcommand{\uu}{\hbox{$\bar{u}u$}}
\newcommand{\dd}{\hbox{$\bar{d}d$}}
\newcommand{\sbs}{\hbox{$\bar{s}s$}}
\newcommand{\pp}{\hbox{$\bar{p}p$}}
\newcommand{\nn}{\hbox{$\bar{n}n$}}
\newcommand{\NN}{\hbox{$\bar{N}N$}}
\newcommand{\DD}{\hbox{$\Delta\bar\Delta$}}

\newcommand{\beq}{\begin{equation}}
\newcommand{\eeq}{\end{equation}}
\def\eqref#1{(\ref{#1})}
\def\naive{na\"{\i}ve}

\begin{document}
\begin{flushright}
{hep-ph/0202234} \\
TAUP--2679-01
\end{flushright}
\vskip1.5cm
\begin{center}
{\Large\bf QCD  and $e^+e^-\to$ Baryon + anti-Baryon}
\end{center}
\medskip
\begin{center}
{\bf Marek Karliner}\footnote{\tt marek@proton.tau.ac.il}
\\
and 
\\
{\bf Shmuel Nussinov}\footnote[7]{\tt nussinov@post.tau.ac.il}
\end{center}
\vskip1cm
\begin{center}
School of Physics and Astronomy\\
Raymond and Beverly Sackler Faculty of Exact Sciences\\
Tel Aviv University, Tel Aviv, Israel\\
\end{center}
\begin{abstract}
We discuss the QCD constraints on
$\ee\to$ baryon--anti-baryon close to threshold, in light of the  
puzzling experimental data which indicate that 
close to threshold
 $\sigma (e^+e^-\to n \bar n ) > \sigma (e^+e^-\to p \bar p )$.
We focus on the process $\ee\to\DD$, which is particularly simple
from the theoretical point of view. In this case
it is possible to make exact QCD predictions 
for the relative yields of the four members of the 
$\Delta$ multiplet, modulo one crucial dynamical assumption.
\end{abstract}
\vfill
\eject

 It has been pointed out recently
\cite{Karliner:2001ut},\cite{Ellis:2001xc}
that the
 experimental results
\cite{Antonelli:1998fv}-\cite{Bardin:1994am}
indicating that
\beq
\sigma (e^+e^-\to n \bar n ) >
\sigma (e^+e^-\to p \bar p ) 
\label{nnVSpp}
\eeq
  over the threshold region, $W = 1.88\div 2.45$ GeV, are rather surprising.

  Indeed, if the fragmentation amplitude 
$F(u\to\hbox{proton}) =F(d\to\hbox{neutron})$
  is not smaller in magnitude than
$F(d\to\hbox{proton}) =F(u\to\hbox{neutron})$,
  then the \pp\ cross section is larger than that for \nn.
  This is so since the amplitude for  \uu\ pair production
  by the virtual photon is enhanced by the ratio of quark charges
  $|Q_u/Q_d|=2$,
relative to the corresponding \dd\ amplitude.

  It has therefore been argued 
\cite{Karliner:2001ut},\cite{Ellis:2001xc}
that such reasoning -- inspired by
  perturbative QCD, does not apply at $N \bar N$ threshold. 
The $q\to B$ and $\bar q \to \bar B$ fragmentation functions are not
independent,
annihilation
  channels into multipion states dominate, and other, say Skyrme-like
descriptions for the nucleons are more appropriate than the
``\naive" three quark picture.

Let us briefly restate the main argument of \cite{Ellis:2001xc}.
  The $I=1$ (say $\rho, \rho^\prime, \rho^{\prime\prime}$, etc.) 
intermediate states leading to an even
number of pions dominate the annihilation into multi-pion states.
 This  {\em is} expected from the direct quark EM couplings
  of the \uu\ and \dd\ quarks to the photon in an ideal ``nonet"
 (or ``quartet" when \sbs\ production is  neglected)
  symmetry of the vector multiplets
   (we would like to emphasize that such
  symmetries arise when purely gluonic $1^{--}$ $I=0$ channels are neglected).

  If further the $I=1$ amplitudes dominate the
 \NN\ \  \,\,$1^{--}$ annihilation
  channel -- as the ratio of  $\pp \to  K^+K^-/K^0 \bar K^0$ 
  suggests, then $I=1$  intermediaries dominate $\NN\to\ee$
  at threshold.

 If 
$|A(I=1)|\gg|A(I=0)|$ then
$\sigma(\ee\to\pp) = \sigma(\ee\to\nn)$.
 Further admixture of some $I=0$ amplitude can yield
 the desired experimentally measured cross section ratio
 for a range of
 the parameters $\epsilon\equiv|A(I{=}0)/A(I{=}1)|$  
and $\hbox{Arg}[A(I{=}0)/A(I{=}1)]$.

 In the following we focus on another baryon-antibaryon threshold
 process which was briefly alluded to in
\cite{Karliner:2001ut},
namely the (pair)
production  of $I=3/2$,  $J=3/2$ \  $\Delta$ states: 
$\Delta^{++},\Delta^{+},\Delta^{0}$ and $\Delta^{-}$.

   Our discussion  does not contradict the above phenomenological
  considerations of the \NN\ threshold process for the 
$I=1/2$ nucleons.
  Yet it suggests a complementary point of view and  possible
  ``exact"  QCD predictions even in this non-perturbative regime,
modulo one crucial assumption, as discussed in detail below.

We start by comparing the two processes,
\beq
\ee \,\,\to\,\, \Delta^{++} \overline{\Delta^{++}}\quad:\qquad
\ee \,\,\to\,\, \gamma^*\,\, \to\,\, u\bar u 
\,\,\to \,\, \underbrace{uuu}_{\Delta^{++}} \,+\,
\underbrace{\bar u \bar u \bar u}_{\overline{\Delta^{++}}}
\label{Dpp}
\eeq
   and
\beq
\ee \,\,\to\,\,\, \Delta^{-}\, \overline{\Delta^{-}}\,\,\quad:\qquad
\ee \,\,\to\,\, \gamma^*\,\, \to
\,\,\, d\bar d
\,\,\to \,\, 
\underbrace{ddd}_{\Delta^-}\,\,
+\,\,\underbrace{\bar d \bar d \bar d}_{\overline{\Delta^-}}
\label{Dnn}
\eeq

  Unlike the $\ee\to\NN$ case, where one has
a mixed-symmetry nucleon with two different
fragmentation functions, only one quark-baryon fragmentation function
is involved here, namely 
$ F(u\to\Delta^{++}) = F(d\to\Delta^-)$\ .
Thus we have a clear
  prediction of the ``\naive" quark model,

\beq
\sigma (\ee\to\Delta^{++} \overline{\Delta^{++}}) = 
4 \sigma (\ee\to\Delta^{-} \overline{\Delta^{-}})
\label{naiveD}
\eeq

  The key point that we wish to emphasize is that this prediction is far
  from \naive. 
Eq.~\eqref{naiveD}
above  and other predicted  ratios of  reactions
  involving $\Delta(1238)$ baryons become exact when the
  following assumptions are made:
\hfill\break
\noindent
(a)
we take the $m_u$ = $m_d$ limit; \
(b)
we work to lowest order in $\alpha_{QED}\,$; \
(c)
we consider only ``flavor connected" processes in which the initial
    $q_i \bar q_i$
produced via the initial (virtual) photon do not
 annihilate into gluons and appear in the final hadrons.

 While (a) and (b) are standard, assumption (c) is not as clear-cut.
It entails a $1/N_c$ approximation \cite{'tHooft:1974jz}
extended to baryons \cite{Witten:1979kh}-\cite{Jenkins:1998wy},
and in the present context it is equivalent
to the famous Zweig (OZI) rule.
\footnote{The extension to baryons of involves some subtleties which will be
discussed in some detail a bit later.}

 The $1/N_c$ factor  suppresses flavor-disconnected  
contributions where the initial external current couples via an internal
quark loop, relative to flavor-connected processes where the initial quarks
produced by the EM current are incorporated into final hadrons.

We wish to stress that the issue of OZI rule applied to baryons 
is quite delicate:
while it works well for mesons, it is well known that in the baryon sector
there are large OZI violations, at least in one case, that of
$\bar s s$ meson production in $B\bar B$ annihilation at rest
\cite{Ellis:1989jw},\cite{Ellis:1995ww}. It is important to see whether
this is compatible with assumption (c), to be discussed in more
detail later.
Consequently, we now proceed on the basis of all three assumptions.
Clearly, confronting the ensuing predictions with
experiment will be a sensitive test of assumption (c).

 First, we note that relation \eqref{naiveD} is valid to 
{\em all} orders in perturbation
 theory.
   For any QCD Feynman diagram contributing to
$\ee\to\Delta^{++}\overline{\Delta^{++}}$, 
there is a corresponding diagram, with all 
$u$-quark propagators replaced by (identical!) $d$-quark propagators,
 which contributes to 
$\ee\to\Delta^{-}\overline{\Delta^{-}}$.
Because of the flavor
 independence of gluonic couplings these two diagrams will therefore be
 identical as far as QCD is concerned. The only difference is that
$Q_d=-1/3$ replaces $Q_u=2/3$ in the coupling to the external photon.
 Thus we expect that the amplitudes for the two different processes will
 satisfy

\beq
A(\ee\to\Delta^{++}\overline{\Delta^{++}} )= 
{-2} A(\ee\to\Delta^{-}\overline{\Delta^{-}})
\label{DeltaAmps}
\eeq

   We omitted all the kinematical variables, namely the  external momenta
  and helicities,  which  should be the same for the two processes.

 Upon  adding all the squares of the helicity amplitudes we obtain the
 differential cross section and \eqref{naiveD} above follows.

 Needless to say, we do {\em not} use here any approximate 
($S$-wave)
 symmetric, \naive, quark model or any other model for the wave- (or
 fragmentation) functions, and  allow for any admixture of orbital/radial
 excitations -- and contribution of states with any number of gluons and/or
 extra $q\bar q$ pairs. The only issue at stake here is the 
$u\leftrightarrow d$ exchange
 symmetry. This is a {\em vectorial} flavor symmetry.  The Vafa-Witten theorem
\cite{Vafa:1984tf}, which relies on rigorous QCD 
inequalities (see \cite{Nussinov:1999sx} and \cite{QCDenc} for reviews)
states that such
 symmetries do not break down spontaneously, and hence the all-order 
 perturbative relations should be true in the full theory.

Our argument is actually {\em independent} of any perturbative expansion.
  All hadronic amplitudes can be obtained by
Fourier transformation  and analytic continuations of
 Euclidean $n$-point correlators of (color singlet) local observables. 

 The relevant correlators for the processes of interest are the 
three-point functions of the form:
\beq
K_{\Delta^{++}}(x,y,z) = 
\langle 0 | J_{EM}(x) \,\Delta^{++}(y) \,\overline{\Delta^{++}}(z)|0\rangle
\label{KDpp}
\eeq
with
$J_{\mu,EM}(x)= 
{2\over3} \bar u(x) \gamma_\mu u(x) -{1\over3} \bar d(x) \gamma_\mu d(x)
+\ldots$ \ 
 the electromagnetic current and 
$\Delta^{++}(x)$ a local operator trilinear in  quark
 fields with the quantum numbers of the  $\Delta^{++}$ particle,
\beq
\Delta^{++}(x) = u_{a,i}(x)\,u_{b,j}(x)\,u_{c,k}(x) \,\epsilon^{abc}
\,\Gamma^{ijk}
\label{Ddef}
\eeq
 with $a,b,c$ $(i,j,k)$ color (spinor) indices. $\epsilon$ 
and $\Gamma$ guarantee coupling to an overall color-singlet 
baryon with  total spin $J=3/2$.
  We have an analogous expression for 
$K_{\Delta^{-}}(x,y,z)$.

  There are two contraction patterns of the fermionic operators in the
above correlators -- the ``flavor connected"  and the ``flavor  disconnected".

 In the first case the initial quark(anti-quark) created  from the vacuum
 via $J_{EM}$ at $x$ propagates to the final $y$
(or $z$) vertex.
 In the second case, these quarks annihilate, forming a closed loop which
 starts and terminates at the same point $x$. The path integrals for
 evaluating these contributions are respectively
\beq
K_{\Delta^{++}}^{connected}(x,y,z) = 
Q_u \int d\mu(A)
\hbox{Tr} \left\{
\gamma_\mu 
S_u^A(x,y)
\epsilon \,\Gamma
S_u^A(z,y)
S_u^A(z,y)
\epsilon \,\Gamma^\dagger
S_u^A(z,x)
\right\}
\label{Kconn}
\eeq
and
\begin{eqnarray}
K_{\Delta^{++}}^{disconn.}(x,y,z) = 
\phantom{aaaaaaaaaaaaaaaaaaaaaaaaaaaaaaaaaaaaaaaaaa}
\vrule height 0ex depth 2.6ex width 0pt
\nonumber\\
\sum_{i=u,d,s} Q_i \int  d\mu(A)\,
\hbox{Tr} \left\{ \gamma_\mu S_i^A(x,x) \right\}\,
\hbox{Tr} \left\{ \epsilon \,\Gamma
S_u^A(z,y)
S_u^A(z,y)
S_u^A(z,y)
\epsilon \,\Gamma^\dagger
\right\}
\label{Kdisconn}
\end{eqnarray}
  where the above traces are in color/spinor space , 
$S_i^{A} (x,y)$ is the
 propagator from $x$ to $y$ of the $i$-th flavor quark in a given 
background gauge field $A_\mu(x)$, and
\beq
d \mu(A)=
{\cal D} A_\mu(x) 
e^{-\int d^4x \left({\bf E}^2 +{\bf B}^2\right)}\:
\raise0.08em\hbox{$\displaystyle\mathop{\Pi}_i$}\,
\hbox{Det} (D\!\!\!\!\slash\,_A + m_i)
\label{Ameasure}
\eeq
is the measure in the path integral.

The two contraction patterns are illustrated schematically in Fig. 1(a)
 and 1(b), respectively. 
\begin{figure}[h]
\bigskip
\centerline{\epsfig{file=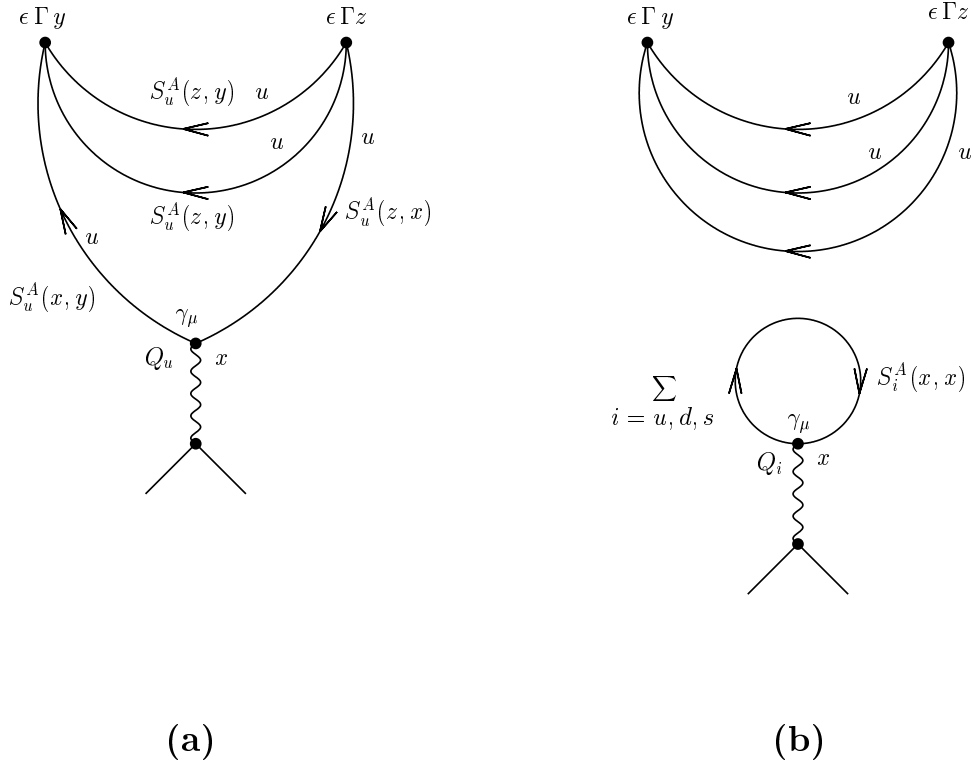,width=12.8cm,angle=0}}
\vspace{0.2in}
\caption{\it \small The two contraction patterns for the fermionic operators
determining the three-point point functions 
$\langle J_{EM} \Delta\bar\Delta\rangle$
in eq.~\eqref{KDpp}:
(a) -- the flavor-connected contraction,
eq.~\eqref{Kconn};
(b) -- the flavor-disconnected contraction, 
eq.~\eqref{Kdisconn}. The quark loop in (b) includes a sum
of contributions from all the light flavors.}
\label{fig:fig1}
\end{figure}

Baryon anti-baryon production is expected to be strongly suppressed
in the large-$N_c$ limit.
Still, it is meaningful to {\em compare} the $N_c$ hierarchy of the two
classes of processes -- 1(a) and 1(b).

Apart from an external color-singlet insertion, 1(a) is essentially the same
(in $N_c$ terms) as the baryonic propagator. The contribution 1(b), on the
other hand, has a quark loop in addition to the baryonic propagator.
As is well known from the double(single) line counting rules for 
gluonic(quark) 
propagators \cite{'tHooft:1974jz}, such a quark loop introduces an
extra $g^2$ factor, yet it does {\em not} increase the number of color
traces. 

Hence the 1(b) flavor-disconnected contraction is suppressed by
$1/N_c$ relative to the connected contraction 1(a).
 Analogous expressions can be written for the correlator
$K_{\Delta^{-}}(x,y,z)$ and
 for the flavor-connected and flavor disconnected-parts thereof.

This argument is quite simple for mesons, but the 
extension to baryons,
although yielding the same qualitative conclusion
\cite{Carone:1993dz}-\cite{Jenkins:1998wy}, involves some subtleties.

Although the result is not new, for the reader's convenience
we review the argument in the following.
Consider the large-$N_c$ baryon propagator with the leading quark-loop
contribution.  It is the diagram with a quark-loop correction to a gluon
propagating between two valence quarks. There are two vertices where
the gluon couples to the valence quarks and two vertices where the quark
loop couples to the gluon.  Each vertex is associated with 
$1/\sqrt{N_c}$, and thus each individual diagram of this kind is 
suppressed by $1/N_c^2$
with respect to the diagram with only valence quarks propagating.

The subtlety arises because there are $N_c(N_c-1)/2$ such pairs of valence
quarks in the baryon, and so the total contribution of the diagrams
with a quark loop is of order ${\cal O}(1)$ and thus appears to 
be unsuppressed.

However, as stressed by Witten \cite{Witten:1979kh}, 
diagrammatic expansion is 
not a very convenient way to study the large-$N_c$ limit of baryons,
because in the perturbation series each successive order grows as a 
higher power of $N_c$.
Thus the diagram with one gluon exchanged between the two quarks in the
baryon scales like ${\cal O}(N_c)$, 
(two vertices $\sim 1/\sqrt{N_c}$ each, multiplied by
combinatorial factor of $N_c(N_c-1)/2$ of possible quark pairs), a diagram
with two gluons exchanged between two different pairs of quarks is of
order $N_c^2$, etc.
This divergence is simply an an artifact of the baryon mass being 
$\sim {\cal O}(N_c)$.

The crucial point for our discussion is that the quark loop is $1/N_c$
suppressed with respect to the contribution from the same diagram without
the quark loop: thus the diagram with one gluon exchanged between two
valence
quarks scales like ${\cal O}(N_c)$, while the same diagram with an additional
quark loop correction to that gluon scales like ${\cal O}(1)$, etc.

The general arguments given by Witten make it clear that the total
contribution of gluons to the baryon propagator is of the same order
as that of the valence quarks, i.e. linear in $N_c$. Since any diagram
with a quark loop is $1/N_c$ suppressed with respect to the same diagram
with only valence quarks and gluons, it follows that quark loops are
suppressed by $1/N_c$ with respect to the valence quarks
\cite{Carone:1993dz}-\cite{Jenkins:1998wy}.

In the real world,
this suppression manifests itself, in particular via the absence of "exotic"
resonances (involving extra $\bar q q$ pairs)
and the narrowness of the ordinary, non-exotic resonances. These hold
equally well both for mesons and for baryons.

 Apart from the overall quark charge factors of  \ ${2\over3}$ 
and -${1\over3}$,  respectively,
 the connected $K_{\Delta^{++}}(x,y,z)$ and $K_{\Delta^{-}}(x,y,z)$
are identical. This identity
 holds in a very strict sense, namely pointwise in the path integrals
\eqref{Kconn} and \eqref{Kdisconn},
i.e for each gauge field configuration $A_\mu(x)$
separately. Hence it is guaranteed to
 hold also for the integrated quantities. This is true 
in any scheme, such as lattice gauge theory, used in order to define
and regularize the path integral in a rigorous way.

Thus we conclude that
\beq
K_{\Delta^{++}}^{connected}(x,y,z) = {-}2
K_{\Delta^{-}}^{connected}(x,y,z)
\label{KconnEq}
\eeq
 holds for any $x$, $y$ and $z$. It will therefore hold for the
 physical amplitudes for \ee\ annihilations into the corresponding
$\Delta\bar\Delta$
 baryons which are obtained by common manipulations of the two
 Euclidean correlators. Thus, to the extent 
that we neglect the 
flavor-disconnected part, relations \eqref{naiveD} 
and \eqref{DeltaAmps} 
\interfootnotelinepenalty=10000
ensue.\footnote{These predictions are shared by
another frequently used nonperturbative approach.
This is the model utilizing $q\bar q$
pair production via tunneling in a chromoelectric flux tube 
\cite{Casher:1979wy}, incorporated in the Lund model \cite{Lund}.
Flavor-disconnected parts are neglected in such models.}

It is important to note that for the case at hand the $K^{disconn.}$
contribution is
suppressed not only by the $1/N_c$\,.
There is also an independent effect which works in the same direction,
namely a partial cancellation of the contributions of the three light quark
 flavors $u$, $s$ and $d$. The cancellation, due to 
$Q_u+Q_d+Q_s=0$, is {\em exact} in
 the $SU(3)$ flavor symmetry
 limit. The physical world is quite well approximated by this limit, since
the $m_s - m_{u/d}$\, mass difference is negligible in comparison with
the mass scale relevant for the 
problem $\sim m_N = 1$ GeV. 
In this case the three terms, due to $u$, $d$ and $s$ loops appearing in
$K^{disconn.}$ above are identical as far as QCD is 
concerned.\footnote{It should be emphasized that unlike
the ``universal" $1/N_c$ suppression, this charge-cancellation mechanism
is relevant only in the specific energy range considered here. At much
lower energies only \uu\ and \dd\ contribute, and at much higher energies
$\bar c c$ contribute as well.}

   The  above arguments apply also for the case of annihilations into
\NN\ states. However, in the latter case we have, even when only the
 flavor-connected parts are retained, {\em two} distinct contributions.
 One  contribution occurs when
 the primary quark produced by the photon is one of the two same-flavour
 quarks in the  nucleon (namely the $u$ in the proton or the $d$ in the 
 of the neutron). The other contribution corresponds to the case when the
 other quark ($d$ for proton or $u$ for the neutron) 
is the primary one.
There is no {\em a priori} model-independent relation between these
 two contributions, nor do we know of any systematic expansion like 
$1/N_c$ which could help relate them, and hence no analogous firm
 prediction can be made here.

In passing we note that also
\beq
K_{\Delta^{++}}^{disconn.}(x,y,z) = 
K_{\Delta^{-}}^{disconn.}(x,y,z)
\label{KdisconnEq}
\eeq
 and the resulting flavor-disconnected
contributions to the physical amplitudes are also
 identical. This is indeed expected, as the disconnected amplitudes
 correspond to purely gluonic and hence $I=0$ intermediate states, which
 couple equally to all members of the $\Delta(1238)$ 
multiplet.

  Considering next  the $I=1$ and $I=0$ isospin
 channels  we  find that eq.~\eqref{KconnEq} above implies that for the
$\gamma^* \to \Delta\bar\Delta$,
\beq
 A (I=1) =2A (I=0).
\label{gDD}
\eeq

Interestingly,
this is consistent with the dominance of the $I=1$ channel 
in $\ee\to\NN$, as inferred from  phenomenological analysis
of the data in Ref.~\cite{Ellis:2001xc}.
It is reassuring that here it follows directly from QCD, 
augmented by assumption $c$.

Isospin symmetry and the above amplitude ratio imply
\beq
 A(\Delta^{++}) : A(\Delta^+) : A(\Delta^0) : A(\Delta^-) = 
2:1:0:{-}1
\label{DampRatio}
\eeq

Thus 
the $\ee\to\Delta\bar\Delta$ cross section ratios 
4:1:0:1 discussed in \cite{Karliner:2001ut}, are in fact exact QCD results, 
modulo assumption (c) above, i.e. 
neglect of the ``flavor disconnected" 
parts.\footnote{For the  processes 
$\ee\to N\bar \Delta$,
involving nucleon and delta production, only the $I=1$ amplitude
contributes and the  ratios are fixed by isospin alone.}

It is interesting to consider the effect of possible resonances in the
$I=1$ ($\rho$, $\rho^\prime$, $\rho^{\prime\prime}$) 
and the $I=0$ channel 
($\omega$, $\omega^\prime$, $\omega^{\prime\prime}$).
Clearly, if at any particular energy an $I=0$ or an $I=1$ resonance
dominates, the above ratio of 
\hbox{$A(I=1)/ A(I=0)$}
cannot be maintained.

 We note however that precisely in the large $N_c$  limit, when the
quark annihilation diagrams are neglected, the $I=1$ and $I=0$ states
are degenerate: 
$m_\omega=m_\rho$,
$m_{\omega^\prime}=m_{\rho^\prime}$, etc.
 In reality, it is much easier to maintain the \hbox{$A(I=1)/ A(I=0)$}
ratio, since
the resonances in the $W> 2$ GeV region are broad
 and overlapping. 

Independently of the ordinary $\rho^{(n)}$ and $\omega^{(n)}$ 
resonances and/or annihilation channels, 
dynamical enhancement of the gluonic $I=0$ intermediate state could result 
if a relatively narrow  glueball state with a vector 
$1^{--}$ quantum numbers
 occurs somewhat above the $\Delta(1238)\bar\Delta(1238)$ threshold.
 It would manifest itself via a  transient deviation from \eqref{DeltaAmps}
 when the resonant energy is traversed.

Qualitative arguments \cite{Nussinov:1999sx} suggest that 
the lightest $1^{--}$ state in the
``three-gluon" sector is at least 50\,\% heavier than the lightest
scalar glueball in the ``two-gluon" sector,
 with \ $m_{0^{++}}\simeq 1.6 \,\div\, 1.7$ GeV. \ In other words,
$m_{1^{--}} \gsim 2m_{\Delta}$.
Lattice calculations \cite{Morningstar:1999rf}
indicate a much larger value,
$m_{1^{--}} \sim 3.8$ GeV. 
If this is indeed the case,
the $1^{--}$ glueball is too far above the $\Delta\bar\Delta$
threshold. It would be very interesting to settle this issue.

Following the suggestion of Refs.~\cite{Karliner:2001ut} and
\cite{Ellis:2001xc},
we next consider the $\gamma \gamma \to B \bar B$  
processes.  We again compare $\Delta^{++}$ and $\Delta^-$ pair
productions.
If we neglect the disconnected part, then the same arguments 
imply a ratio of 4 of the flavor-connected
 amplitudes and 16(!) for the corresponding cross sections,
as discussed briefly in \cite{Karliner:2001ut}.

It should be emphasized, however, that the neglect of the 
flavor-\break
disconnected part relies here, unlike for the previous 
single-photon case, on the $1/N_c$ suppression only and {\em not}
on cancellation of charges as  well.
There is no such cancellation  here, as only the squares of
the  charges enter.

In addition, in the two photon case we have also the 
$0^-$ $\eta^\prime$ anomalous channel,
which implies that the flavor-disconnected parts may be enhanced.
However, as we move even slightly above the $B \bar B$ threshold, the number
of partial waves contributing increases rapidly and the relative importance
of the pseudoscalar channel decreases.

The present work is indirectly triggered by present data on $\ee\to\nn/\pp$
at threshold, which seem to confound the \naive\ QCD expectations. 
Our main point is to note that there is a similar process, 
in which the theoretical analysis is much more
straightforward, namely  $\ee\to\DD$. In this case
QCD, augmented by a well-motivated dynamical assumption, 
makes clearcut and largely model-independent predictions.
Future experimental tests of these predictions 
will be a sensitive probe of this dynamical assumption.

\section*{Acknowledgments}
The research of one of us (M.K.) was supported in part by a grant from the
United States-Israel Binational Science Foundation (BSF), Jerusalem,
Israel. Both M.K. and S.N. were supported in part by the 
 Basic Research Foundation administered by the Israel
Academy of Sciences and Humanities. We would like to thank John Ellis for
useful comments on the manuscript.

\end{document}